# Mathematical modeling of morphological changes in photochromic crystals by catastrophe theory

Hirotsugu Suzui[1], Kazuharu Uchiyama[2], Kingo Uchida[3],
Ryoichi Horisaki[1], Hirokazu Hori[2] and Makoto Naruse[1]

[1] Department of Information Physics and Computing, Graduate School of Information Science and Technology, The University of Tokyo, 7-3-1 Hongo, Bunkyo-ku, Tokyo 113-8656, Japan.

[2] University of Yamanashi, 4-3-11 Takeda, Kofu, Yamanashi 400-8511, Japan.

[3] Department of Materials Chemistry, Ryukoku University, 1-5 Yokotani, Oecho, Seta, Otsu, Shiga 520-2194, Japan.

E-mail: tsuguh-hirosmtronica@g.ecc.u-tokyo.ac.jp

**ABSTRACT**

Photochromic diarylethene is known to exhibit reversible photoisomerization under irradiation with ultraviolet (UV) and visible light. Besides reversible optical properties upon light irradiation, a variety of discontinuous morphological changes are reported in the literature, such as sudden crystal bending, cracking, and photosalient effects, which are caused simply by UV and visible light irradiation. These morphological phenomena with discontinuities are micro-scale changes caused by photoisomerization at the nanoscale and lead to the realization of important functions as optical devices. However, the theoretical models behind these phenomena are not well understood. In this paper, we construct a mathematical model that can treat diverse phenomena in a unified model by using swallow-tail catastrophe, a higher-order catastrophe than cusp catastrophe, from the seven elementary catastrophes that can describe discontinuities in the phenomena. By introducing the hyperbolic operating curves in the model, intrinsic properties of the photochromic crystals are represented. The induced morphogenesis, such as bending, cracking, and photosalient, are systematically classified by the proposed catastrophe model, which even implies unexplored operating conditions of the crystals and explains known phenomena. The proposed catastrophe-theory-based modeling provides a foundation for understanding and discovering the versatile morphogenesis in photochromic crystals. Furthermore, the proposed approach provides a basis for understanding and discovering various morphological changes in photochromic crystals and similar systems.

# I. INTRODUCTION

Photochromic single crystals of diarylethenes (DAEs) are interesting platforms for realizing intelligent functions based on light–matter composite systems because it is possible to store light information in the nanometer-scale material deformation involved in the reversible photoisomerization induced by ultraviolet (UV) and visible light irradiation.[1,2] Under UV light irradiation, DAEs undergo isomerization from an open-ring isomer (referred to as **1o** in Fig. 1(a)) to a closed-ring isomer (**1c**), changing their color from transparent to blue opaque. Conversely, visible light induces isomerization from **1c** to **1o**, inducing a change in color from blue opaque to transparent. DAE exhibits microscopic changes in its molecular structure and volume and macroscopic changes in its absorption spectra (Figs. 1(a) and 1(b)).[3,4]

Besides such changes in color, photochromic crystals are known to show various morphological changes depending on their molecular structures induced by light irradiation. Some crystals show so-called photosalient phenomena, in which the crystals autonomously break due to molecular distortion caused by UV light irradiation.[5,6] Meanwhile, the formation of subwavelength-scale, autonomous groove structures has been reported, which is another example of a morphological change in photochromic crystal systems.[7] Light-induced bending of photochromic crystal is another notable demonstration where the crystal undergoes a gradual bending away from the light source (Figs. 1(c,i) and 1(c,ii)), followed by a sudden bending toward the light source (the reverse direction) (Figs. 1(c,iii) and 1(c,iv)), and then cracks (Fig. 1(c,v)).[8] In this paper, we refer to this phenomenon as bending-and-cracking phenomena. These phenomena are macroscopic changes that result from microscopic changes, leading to new functionalities.

Some phenomena in photochromic crystals exhibit *discontinuous* changes. In the case of the bending-and-cracking phenomenon, Fujimoto *et al.* provided a movie of the sudden bending of photochromic crystal in the Supplementary Material of Ref. [8]. We analyzed the movie via image processing to quantitatively examine the sudden structural deformation. Figure 1(d) characterizes the horizontal position of the tip of the crystal relative to the initial position. The spatial position is extracted via signal processing. Until approximately 2.7 seconds, the horizontal position increase toward the positive side which is indicated away from the light source (toward the right side in the movie), as indicated by Fig. 1(c,i). However, at the time indicated by Fig. 1(c,ii), the horizontal position decrease toward the negative side which is indicated toward the light source (left in the movie), leading to an abrupt change in the horizontally negative position, as shown in Fig. 1(c,iii). An experimental demonstration of cracking after bending is appended as a movie in the Supplementary Material. Image processing analysis of this movie is also presented in the Supplementary Material.

Including the above-demonstrated bending-and-cracking phrnomena, a variety of discontinuous morphology changes in photochromic crystals are found in the literature. However, their underlying theoretical structure is not well understood. One of the important aspects is that both crystal and light properties must be taken into account, and sudden or discontinuous changes must be represented.

To resolve these issues, here we propose a theoretical model based on catastrophe theory for a unified understanding of unique morphological changes observed on the macro scale that are caused by changes on the microscopic scale in photochromic crystals. Catastrophe theory, which is reviewed briefly below, provides a rigorous foundation for morphology changes, including discontinuous changes. By considering the photoisomerization properties in photochromic crystals in catastrophe-based modeling, discontinuous morphological changes can be systematically understood. Furthermore, catastrophe-based modeling suggests potentially achievable phenomena or unexplored operating scenarios of photochromic crystals, which is another insight gained by the proposed model theory.

The rest of the paper consists of the following. Section II gives a preliminary description of the present study by reviewing catastrophe theory, as well as by introducing the principle theoretical element, which is swallow-tail catastrophe. Section III describes the proposed theoretical model. Utilizing the model, Sec. IV analyzes morphological changes with a variety of parameter variations. Section V concludes the paper.

## II. PRELIMINARY

Catastrophe theory, proposed by René Thom in the 1970s, examines the qualitative change process of discontinuous phenomena from the viewpoint of structural stability by applying differential geometry and topology in a precise mathematical model.[9] The remarkable outcome of Thom's catastrophe theory is that discontinuous phenomena are described using topological images of the phenomena and exact mathematical models, and the phenomena are classified into one of seven categories.[9] Catastrophe theory is widely utilized in diverse disciplines, such as brain modeling in biology,[10] optimization of road transportation systems in sociology,[11] prediction of market price fluctuations in economics,[12] among others.

One remark here is that most of the studies using catastrophe theory to date were performed on the basis of fold catastrophes or cusp catastrophes, which are called low-order catastrophes. The potential ability of catastrophe theory, however, covers more complex systems. In the present study, we examine bending-and-cracking phrnomena exhibited by photochromic crystals, which involve two discontinuous changes. In this study, we utilize swallow-tail catastrophe or high-order catastrophe to examine the complex and dynamic morphological changes observed by the bending of photochromic crystals. It should be remarked that microscopic isomerization induced by ultraviolet light irradiation causes diverse macroscopic morphological deformations, and higher-order catastrophe theory is a compatible and essential vehicle that accounts for such complex light–matter systems.

## III. CATASTROPHE THEORETIC MODELING
### A. Discontinuous changes

The mechanism of the photochromic crystal bending-and-cracking phrnomena is schematically summarized in Fig. 2(a). Experimental demonstrations have been reported in the literature.[13] A material-science-based interpretation is given below.[14,15]

(i) Consider that the photochromic crystal is in a completely open-ring state initially (Fig. 2(a,i)). UV light irradiation from the lefthand-side of the crystal causes isomerization **1o** to **1c** near the surface of the light-irradiated side.

(ii) The change in molecular length accompanying the photoisomerization causes an expansion force on the crystal surface, and the entire crystal deflects in a direction opposite to that of the irradiation, indicated by the red arrows in Fig. 2(a,ii).

(iii) However, when a certain amount of surface molecules undergo photoisomerization, the distortion of the entire crystal is increased by the pressure due to the change of the molecular length. To resolve the distortion, the molecular package of the molecules around the isomerized crystal undergoes a phase transition from an open-ring one to a closed-ring one (Fig. 2(a,iii)). Here, the phase transition is defined as a change in the crystal package.[8,17]

(iv) As a result, the volume at the irradiated surface decreases rapidly, and the crystal immediately bends in the opposite direction from the UV light irradiation, denoted by blue arrows in Fig. 2(a,iv).

(v) When crystal in this state is further irradiated with UV light, the crystal cracks (Fig. 2(a,v)). This is considered to be caused by the accumulation of distortion within the crystal.[8] Irradiating a bent crystal (Fig. 2(a, iv)) state with visible light from the same direction as the UV light restores it to a straight state, but changes in the a, b, and c crystal axes compared to the initial state, as observed by X-ray analysis, have been reported in the literature.[8] This is due to the mismatch caused by the package change within the crystal.

Here, first we focus on the crystal distortion and mismatch, which are involved in (iv) and (v) above, independently and adapt cusp catastrophe theory to each of them. The cusp catastrophe has two stable states. The potential equation for the cusp catastrophe is given by

$$V_{uv}(x) = \frac{1}{4}x^4 + \frac{u}{2}x^2 + vx, \tag{1}$$

where $x$ means the state variable, and $u$ and $v$ are the control parameters of the system, whose changes determine the functional state of the system.[9]

The equilibrium space $M_V$ is defined by the space where the derivative of the potential is zero, which is expressed by

$$M_V = \left\{(u, v, x) \mid \frac{\partial V}{\partial x} = x^3 + ux + v = 0\right\}. \tag{2}$$

The singularity set $\Sigma_V$ of the catastrophe map of the potential function is defined by

$$\Sigma_V = \left\{ (u,v,x) \middle| \frac{\partial V}{\partial x} = x^3 + ux + v = 0, \frac{\partial^2 V}{\partial x^2} = 3x^2 - u = 0 \right\}, \quad (3)$$

meaning that the second derivative of the potential is also zero. From Eq. (3), the branch set $B_V$, which is the singularity sets projected onto the $(u, v)$-plane, is expressed as

$$B_V = \{(u,v) | 4u^3 + 27v^2 = 0\}. \quad (4)$$

See the Supplementary Material for the derivation of Eq. (4). The singularity set $\Sigma_V$ represented by Eq. (3) is shown by the surface at the top of Fig. 2(b), whereas the branch set $B_V$ represented by Eq. (4) is indicated by the blue curve shown at the bottom of Fig. 2(b).

What should be noted here is that two equilibrium states $x$ exist in some regime of $(u, v)$ in the cusp catastrophe. For example, assume that the control variables $(u, v)$ are given by the position $S$ in Fig. 2(b), and the equilibrium state $x$ is on the upper surface. Let $v$ increase while $u$ is fixed, as indicated by the red arrow. Here, at the position $M$, the equilibrium state $x$ jumps from the upper surface to the lower surface; namely, a discontinuous change is induced.

Here we discuss the physical picture of morphological changes observed in photochromic crystals from the viewpoint of catastrophe. First, assume that the state variable $x$ means the degree of bending of a photochromic crystal observed at the macro-scale, whereas the control variable $v$ manifests the ratio of strain induced in the photochromic crystal on the micro scale. When the ratio of strain exceeds a tipping point, the photochromic crystal suddenly bends to the opposite side, as discussed before.

This is clearly represented in the cusp catastrophe shown in Fig. 2(c,i) where a discontinuous jump is observed at the strain ratio indicated by $M$. Conversely, when the strain ratio is reduced, it is possible for crystals in a bending state to return to their original state. Such a discontinuous jump could occur at a different $v$ value at $M'$, which is represented by a dashed arrow in Fig. 2(c,i).

Similarly, suppose the control variable $v$ represents the mismatch in crystal structure due to differences in crystal packages. The state variable $x$ means the state of the crystal. In Fig. 2(c,ii), the state starts with a situation where the crystal is sticking together. As the mismatch, represented by the control variable $v$, increases, the crystal experiences a discontinuous change at the point indicated by $M$, and changes to a state where the crystal is cracked. When the ratio of mismatch decreases, the crystal goes back to the original state, but the point of the discontinuous change would occur at a different $v$ value at $M'$, which is denoted by the dashed arrow in Fig. 2(c,ii). One remark here is that such hysteresis itself has not been clearly observed yet in the experiments. From a theoretical standpoint, the hysteresis can be very small if the operating curve crosses over the branch set near the cusp point.

## B. Swallow-tail catastrophe modeling

Although the above-described cusp catastrophe captures individual discontinuous events, in the actual phenomenon, these two events occur simultaneously. Therefore, in order to accommodate these different types of discontinuities in a unified model, which cannot be represented by a cusp catastrophe, we introduce a higher-order catastrophe called the swallow-tail catastrophe, whose potential is given by

$$V_{uvw}(x) = \frac{1}{5}x^5 - \frac{u}{3}x^3 + \frac{v}{2}x^2 + wx, \tag{5}$$

where $x$ means the state variable, and $u$, $v$, and $w$ are the control variables of the system.[9] Similarly to the discussion above, the equilibrium space $M_V$ is given by the space where the derivative of the potential is zero, which is expressed by

$$M_V = \left\{(u, v, w, x) \middle| \frac{\partial V}{\partial x} = x^4 - ux^2 + vx + w = 0\right\}. \tag{6}$$

The singularity set $\Sigma_V$ of the catastrophe map of the potential function is expressed by

$$\Sigma_V = \left\{(u, v, w, x) \middle| \frac{\partial V}{\partial x} = x^4 - ux^2 + vx + w = 0, \frac{\partial^2 V}{\partial x^2} = 4x^3 - 2ux + v = 0\right\}. \tag{7}$$

From Eq. (7), the branch set $B_V$, which is the singularity sets of the swallow-tail catastrophe projected onto $(u, v, w)$-space, is expressed by

$$B_V = \{(u, v, w) | \exists x, x^4 - ux^2 + vx + w = 0, 4x^3 - 2ux + v = 0\}. \tag{8}$$

The singularity set $\Sigma_V$ and the branch set $B_V$ of the swallow-tail catastrophe at a specific $u$ are shown at the top and bottom of Fig. 2(d), respectively.

There are two remarkable properties inherent in the swallow-tail catastrophe under the condition $u > 0$. One is that there are two cusps or apexes in the branch set, leading to the unified description of different kinds of discontinuous events, which is discussed in detail below.

The other is that the so-called fold catastrophe is present in the swallow-tail catastrophe. In the fold catastrophe, a state cannot exist in a certain control parameter regime. More specifically, the impossible zone of control parameters $(v, w)$ is denoted by a gray-colored area in Fig. 2(e). We discuss extensions regarding the parameter $u$ later below. The photosalient effect, meaning that the crystal autonomously breaks into many pieces, corresponds to such an impossible area of the state variable, which will be discussed in detail later below.

For the unification of two kinds of discontinuous phenomena in photochromic crystals, the aforementioned cusp structure and straight operating curve cannot be used to describe them. Therefore, we introduce a hyperbola for the operating curve, indicated by the solid red curve and dashed red curve in Fig. 2(e), which is in the form

$$w^2 - v^2 = e^2 \tag{9}$$

where $e$ is the eccentricity of the hyperbola. Here, there are two asymptotic lines or asymptotes of the hyperbola, denoted by the orange solid and dashed lines in Fig. 2(e). For simplicity, we assume that the *w*-axis value of the origin coincides with that of the apex of the branch set. In this case, the intersection of the asymptotic lines is given by $(v_0, w_0) = (0, -u^2/12)$. See the Supplementary Material for details of the derivation.

We associate physical meanings with the control parameters *v* and *w*. The *v*-axis is regarded as the photoisomerization rate of the crystal, which is set by adjusting the external light irradiation. That is, the execution of the operating curve from the initial state to the final state corresponds to a monotonic increase of the *v* value.

On the other hand, we consider that the *w*-axis is associated with the softness of the crystal. This is based on the fact that the photochromic crystal becomes soft in a mixed state of the colored and transparent crystalline states during photoisomerization.[16] In the bending-and-cracking phrnomena of the present study, the softness at the initial and the final state is considered to be similar, while the crystal undergoes a state where the crystal becomes the most softened. The *w*-axis of the hyperbolic operating curve accommodates such a property, where *w* takes a minimum value at $v = 0$, whereas the *w*-value of the initial and the final states is the same.

The irradiation of UV light causes the crystal to travel along the solid red curve, exhibiting a sudden bending when the operating crosses over the second branch set marked by M1 in Fig. 2(e). By irradiating the crystal with visible light, the crystal goes back to the initial non-bending state. However, multiple repetitions of irradiation of visible and UV light (that is to say, repetition of bending and resetting) causes accumulated mismatch of the crystal due to multiple-time phase transitions. The operating curve then follows the dashed red curve shown in Fig. 2(e); when the operating curve meets the second crossing over the branch set denoted by M2, the crystal experiences another type of morphological discontinuous change, which is cracking.

The asymptotic lines of the hyperbolic operating curve also represent physical meanings. The positive-slope asymptotic line, denoted by a solid orange line, is related to the strain, which corresponds to the *v*-axis (namely, the ratio of strain) in the cusp catastrophe model in Fig. 2(c,i). The negative-slope asymptotic line, denoted by a dashed line, is related to the mismatch of the crystal, which corresponds to the *v*-axis (namely, the ratio of mismatch) in the cusp catastrophe model in Fig. 2(c,ii).

## IV. ANALYSIS

We can think of a variety of operating curves in the (*v, w*)-plane to represent diverse morphological phenomena observed in photochromic crystals. First we deal with the following two representative cases. Although the actual operating curve in this model consists of two curves, for simplicity, this chapter uses only one operating curve for evaluation.

The first scenario regards the bending-and-cracking phrnomena, denoted by the red-colored operating curve in Fig. 3(a), where the curve travels over two cusps, which we call Scenario 1. The starting point of the operating curve is denoted by (i). With the crystal being irradiated with UV light, photoisomerization progresses, which is manifested by the increase of the *v* value over the hyperbola. When the operating curve crosses the second blue curve, denoted by (iii), the crystal exhibits a sudden bending. Such a discontinuous morphological change is due to the cusp point indicated by X.

After that, further photoisomerization progresses along the operating curve. At the point denoted by (iv), the crystal undergoes cracking. Such a sudden morphological change is due to the cusp point indicated by Y. What should be remarked on is that these two consecutive discontinuous changes are represented in a unified manner by this proposed swallow-tail catastrophe-based model.

Another representative scenario is shown by the green-colored operating curve in Fig. 3(a) where the intersection of the asymptotic lines is given by $(v_o, w_0) = (0, -u^2/12 + 7)$. As discussed in Sec. III, the *w*-axis is related to the softness/hardness of the crystal. The green curve stays on the lower side in Fig. 3(a), meaning that the crystal is hard. Indeed, such a hard photochromic crystal has been reported in the literature,[6] and such a scenario is called Scenario 2.

The green hyperbolic operating curve begins with (i). As the photoisomerization progresses, the operating curve intersects with the blue-colored branch set, indicated by (C). It should be noted that the point (C) belongs to the fold catastrophe, as discussed in Sec. III. That is, no state is allowed beyond the point (C), which is shaded in gray. Therefore, when the operating curve reaches the point (C), a discontinuous change occurs, and the crystal breaks into pieces, which is known as the photosalient phenomenon.[6] In other words, once the operating curve meets the fold catastrophe, the broken crystal cannot go back to its original state. Therefore, the green operating curve is terminated at the point (C). To account for such termination, we consider that the lower right portion of the branch set, indicated by the dashed green line, does not exist in counting the number of intersections between the branch set and the operating curve in the following characterization to quantify the fold catastrophe.

In the above analysis, the control parameter *u* was fixed at 5. The branch set on the (*v, w*)-plane, denoted by the blue curve, exhibits a quite different profile depending on the *u* value, as represented *u* = 2, 5, 7, and 10 in Fig. 3(b,i).

Furthermore, the operating curve is characterized by the eccentricity of the hyperbola, denoted by *e*. In Fig. 3(b,ii), three hyperbolic operating curves are shown by the solid, dashed, and dotted lines, whose eccentricities are 1, 5, and 10, respectively, while *u* is 5. The number of intersections between the branch set and the operating curve characterizes the resulting discontinuous phenomena.

We first consider the case where the intersection of the asymptotic lines is given by $(v_0, w_0) = (0, -u^2/12)$, which is Scenario 1 in Fig. 3(a). The number of intersections is classified as 0 to 4, as summarized by the diagram of *u* and *e* values shown in Fig. 4(a). The red-colored region indicates the area with four intersections, meaning that sudden bending followed by cracking is induced. Meanwhile, the dark-gray-colored region shows the area with only one intersection.

From Fig. 4(a), we can observe several characteristics. The red-colored region cannot exist when $u$ is smaller than approximately 4.8 regardless of the $e$ value, which is consistent with the experimental results in Ref. 8. Then, we examine what happens on the line $u = 5$, which implies room-temperature operation. The red area, where the sudden-bending-and-cracking phenomenon occurs, exists only when the $e$ value is smaller than approximately 1. This means that the thickness of the crystal should be small enough. At the same time, when $e$ is greater than approximately 8, the dark gray arises, meaning that the photosalient phenomenon happens even with a thick crystal. Such a photosalient phenomenon has not yet been examined experimentally in the literature. An experimental investigation will be an interesting future study, stimulated by such catastrophe-theory-based modeling analysis.

Second, we examine the case where the intersection of the asymptotic lines is given by $(v_0, w_0) = (0, -u^2/12 + 7)$. The number of intersections between the branch set and the operating curve is summarized in Fig. 4(b). Focusing on the line $u = 5$, the number of intersections is always 1 regardless of the $e$ value, meaning that the crystal exhibits only the photosalient phenomenon.

Moreover, we generalize the model to take further properties into consideration. As discussed already for the red operating curve in Fig. 3(a), the point when the operating curve reaches its maximum along the $w$ axis means the state where the crystal becomes the softest. We call this point the melting point. In photochromic crystals, it is experimentally known that the melting point differs depending on the crystal.[16] That is, some crystals undergo melting with slight photoisomerization. In other words, the melting occurs with a small $v$ value.

Such characteristics can be taken into account by horizontally shifting the asymptotic lines, as shown in Fig. 4(c,i), where the $v$ coordinate of the intersection of the shifted asymptotic lines is specified by $v_o = -5$, whereas that of the original one is specified by $v_o = 0$.

What is of interest is how the number of intersections between the branch set and the hyperbolic operating curve behaves. The diagram in Fig. 4(c,ii) summarizes the number of intersections in a parameter space configured by $u$ and the horizontal shift of the asymptotic lines represented by $v_0$. A small $v_0$ value indicates that melting is induced with slight photoisomerization by UV irradiation, whereas a large $v_0$ value means that melting occurs after sufficient photoisomerization by UV irradiation. Focusing on the $v_0$ dependency on the line $u = 5$, the number of intersections is no longer four when $v_0$ is larger than 5 or smaller than −5, meaning that crystals with extremely high and low melting points no longer cause the bending-and-cracking phenomenon. At the same time, we can argue that the red four-intersection zone is larger than in the case of the eccentricity-dependency analysis in Fig. 3(a). Moreover, the diagram is symmetric with respect to the $v_o = 0$ line.

The asymptotic line can also be shifted along the *w*-axis direction (Fig. 4(d,i)). The vertical translation of the operating curve means that the hardness of the crystal in its initial state is changed. Figure 4(d,ii) summarizes the number of intersections in a parameter space configured by $u$ and the vertical shift of the asymptotic lines represented by $w_0$. Indeed, Scenario 1, represented by the red operating curve in Fig. 3(a), corresponds to a case in the red region with a low but not too low $w_0$ value when $u$ is 5. Scenario 2, shown by the green operating curve in Fig. 3(a), corresponds to a case in the gray zone with a large but not too large $w_0$ value when $u$ is 5. What is revealed in Fig. 4(d,ii) is that the number of intersections evolves from 0, 4, 2, and 1 as the $w_0$ value increases from −10 to 10 along the line $u = 5$. That is, diverse phenomena are actually possible, not just Scenarios 1 and 2.

Additionally, when we pay attention to a specific positive $w_0$ and examine the changes as the *u* value increases, which is indicated by the dotted line in Fig. 4(d,ii), the sudden-bending-and-cracking phenomenon, represented by the red zone, happens with a limited range of *u* values. That is, a properly configured temperature is indispensable for inducing the phenomena. The experimental examination of temperature dependence is an interesting future topic, triggered by catastrophe-theory-based modeling. Through such analysis, the proposed catastrophe model accommodates and systematically understands not only photochromic crystals but also various phenomena with similar systems.

## V. CONCLUSION

On the basis of swallow-tail catastrophe, we examined discontinuous morphological macroscale changes induced in photochromic cs, such as bending and photosalient caused by microscopic photoisomerization. We found that not only the bending-and-cracking phenomenon but also irreversible crystal fracturing and photosalient phenomena can be described by the same theoretical model based on catastrophe theory. That is, a variety of approaches leading to a universal model were demonstrated. A hyperbolic operating curve allows thoughtful and systematic analysis regarding the intersection of the operating curve and the branch set. It should also be emphasized that the agreement between the experimental results and the model theory is not the only advantage of the proposed approach. The catastrophe approach suggests the possibility of designing unknown scenarios of morphogenesis in photochromism. In other words, it reveals unexplored areas of operation and even suggests interesting directions in molecular design.

## SUPPLEMENTARY MATERIAL

See supplementary material for the theoretical derivations.

**Acknowledgments**

This work was supported in part by JST SPRING (JPMJSP2108) and CREST project (JPMJCR17N2) funded by the Japan Science and Technology Agency and Grants-in-Aid for Scientific Research (A) (JP20H00233) and Transformative Research Areas (A) (22H05197) funded by the Japan Society for the Promotion of Science.

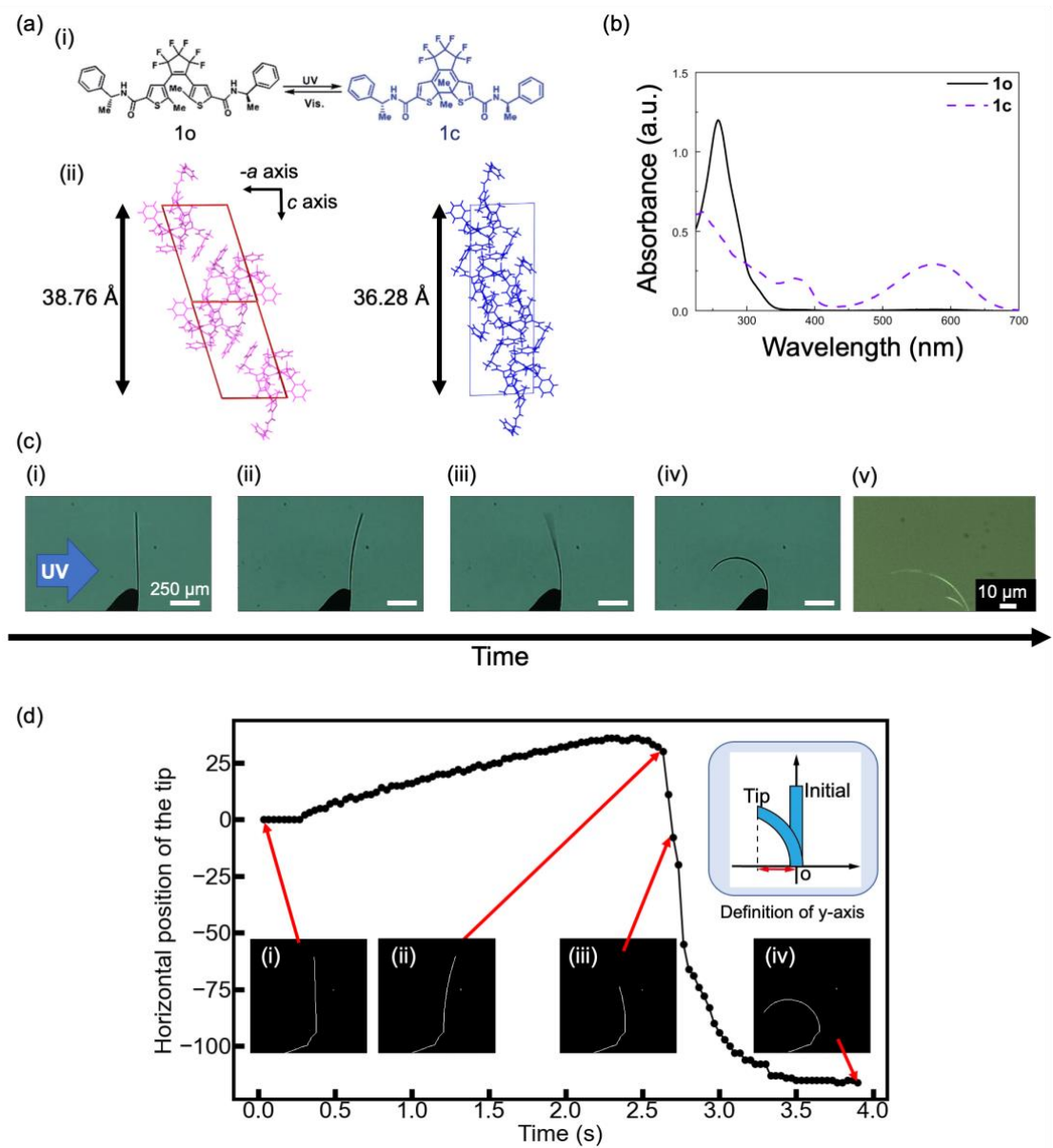

**Fig. 1** (a) (i)The diarylethene molecules used in this study and the photoisomerization (open-ring isomer **1o** and closed-ring isomer **1c**). (ii) The diarylethene molecule packages used in this study [Adapted from A. Fujimoto, N. Fujinaga, R. Nishimura, E. Hatano, L. Kono, A. Nagai, A. Sekine, Y. Hattori, Y. Kojima, N. Yasuda, M. Morimoto, S. Yokojima, S. Nakamura, B. L. Feringa, and K. Uchida, Chem. Sci. (2020). Copyright 2020 The Royal Society of Chemistry][8] (b) Absorption spectra of **1o** (solid black line) and **1c** (blue dashed line). (c) Crystal bending and cracking phenomena are caused by radiating ultraviolet light from one direction. [Adapted from A. Fujimoto, N. Fujinaga, R. Nishimura, E. music Hatano, L. Kono, A. Nagai, A. Sekine, Y. Hattori, Y. Kojima, N. Yasuda, M. Morimoto, S. Yokojima, S. Nakamura, B. L. Feringa, and K. Uchida, Chem. Sci. (2020). Copyright 2020 The Royal Society of Chemistry][8] (d) Results of the horizontal position of the tip analysis of crystal bending phenomena. A discontinuous change is induced after the time indicated by (ii).

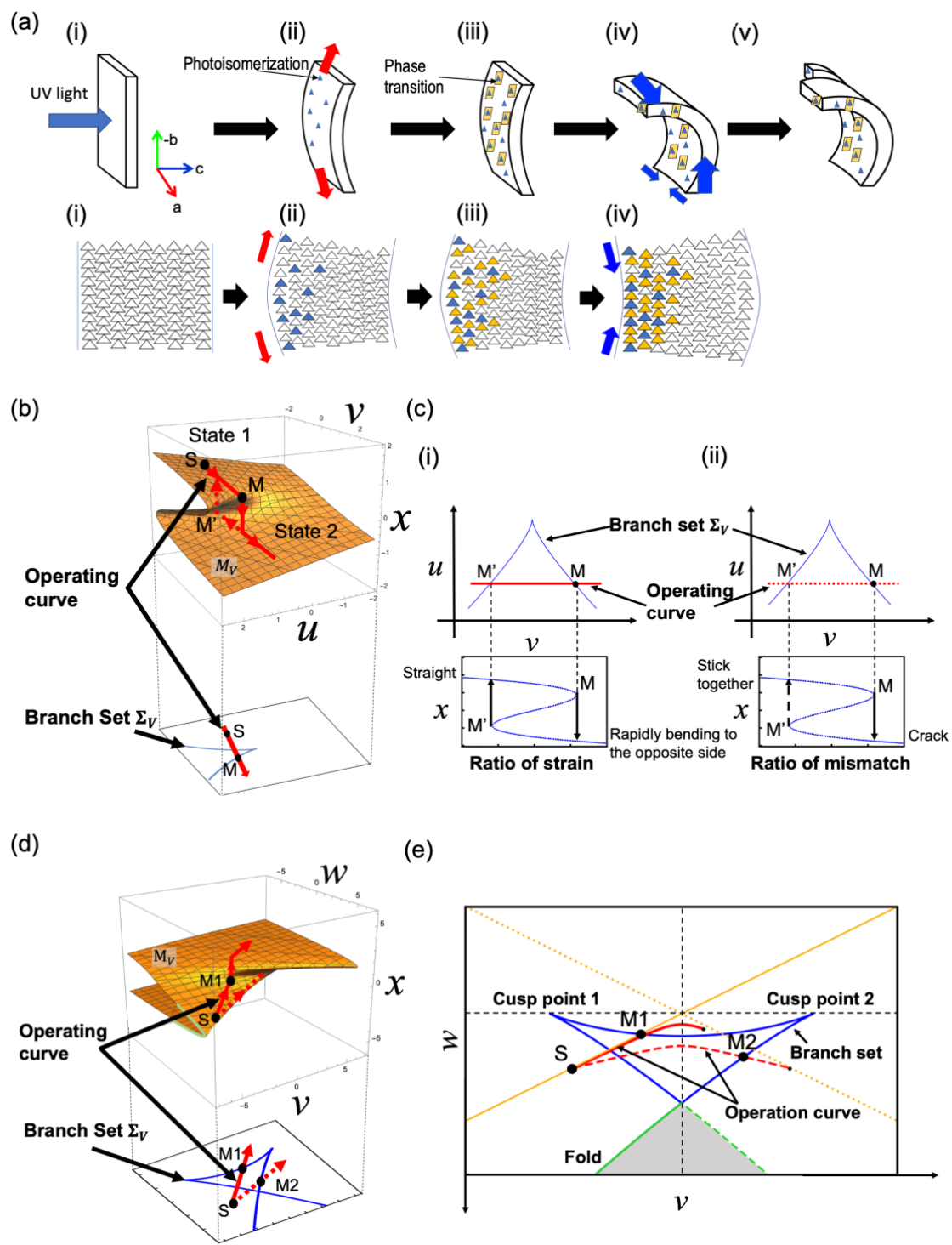

**Fig. 2** (a) Modeling of crystal bending phenomena. (b) Cusp catastrophe potential and branch sets. (c) Elements of bending phenomenon with cusp catastrophe and straight operating curves. (d) Swallow-tail catastrophe potential and branch sets at $u = 5$. (e) Catastrophe and hyperbolic operating curve for catastrophe parameter $u = 5$.

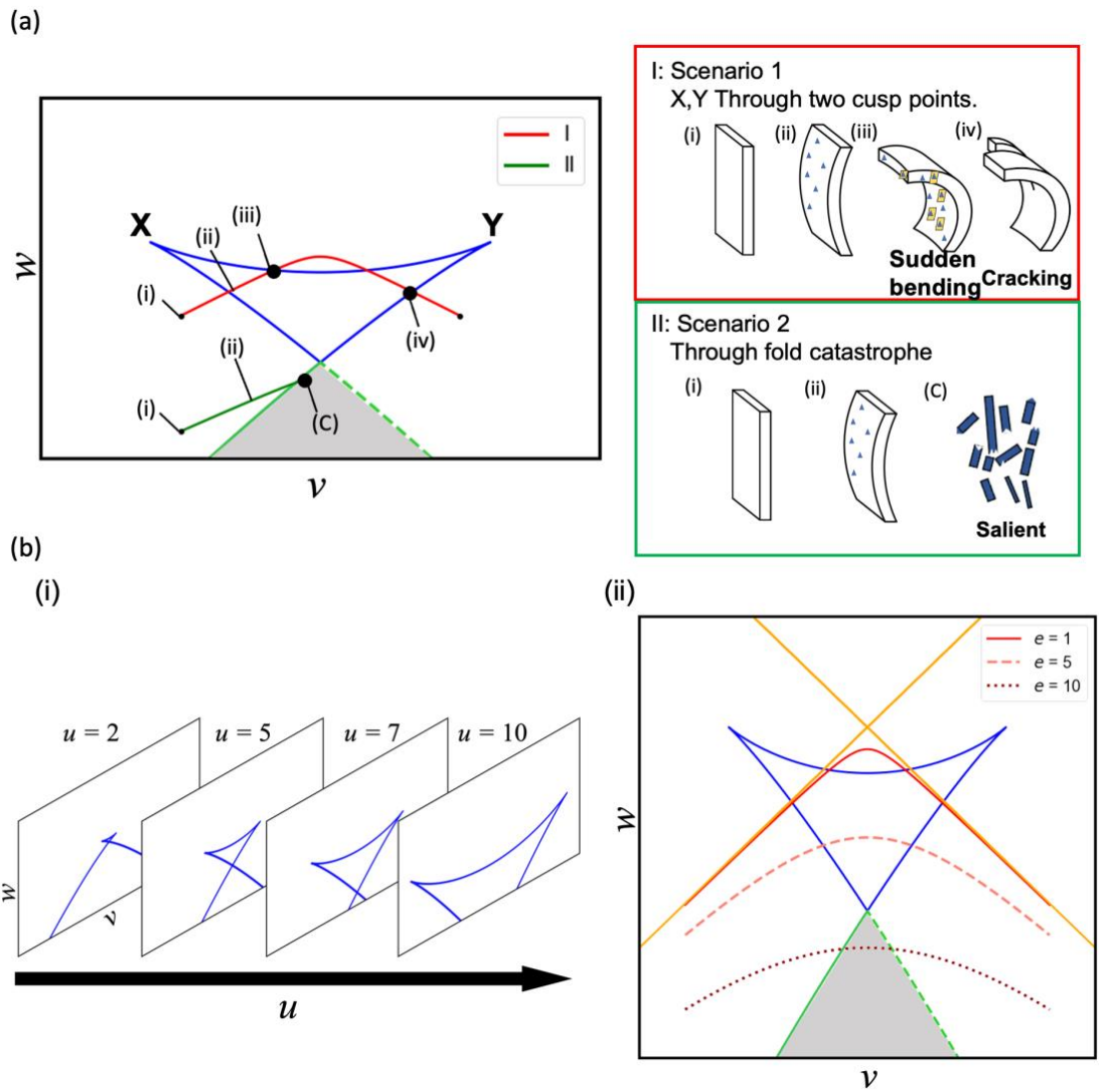

**Fig. 3** (a) Swallowtail catastrophe and different crystal operating curves. (b) (i) Variation of the shape of the branch set depending on parameter $u = 2, 5, 7$, and $10$ from left to right. (ii) Variation of the operating curves specified by the ellipticity of the hyperbola denoted by $e$. The solid, dashed, and dotted curves are specified by $e = 1, 5$, and $10$, respectively.

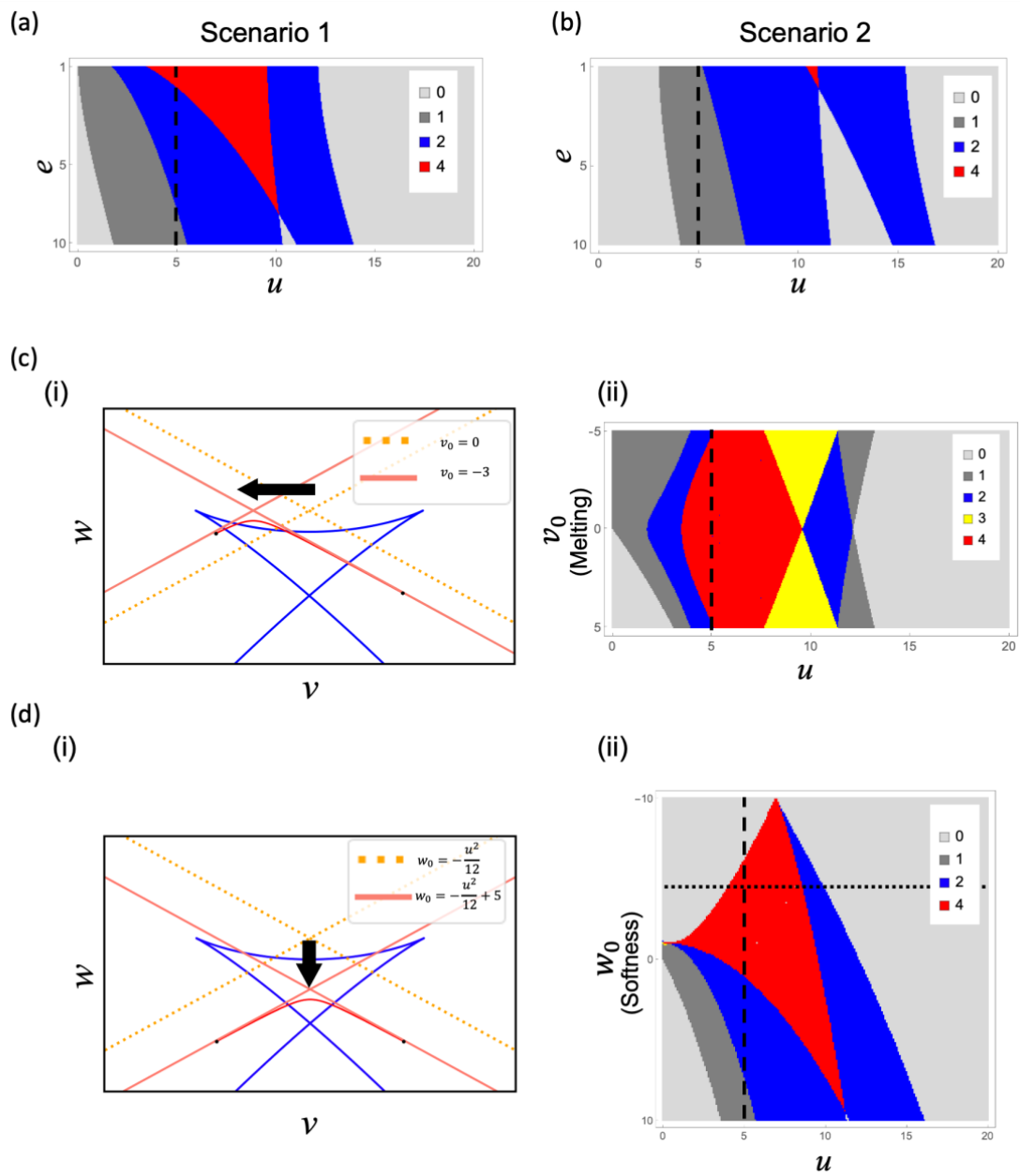

**Fig. 4** (a) Analysis results with Scenario 1. (b) Analysis results with Scenario 2. (c) (i) Model representation of the melting point dependence of crystals by the horizontal translation of the asymptotic lines of the hyperbolic operating curve. The asymptotic lines denoted by the solid lines mean a lower melting point than the dotted ones. (ii) Results of analysis at temperature and crystal melting points. (d) (i) Model representation of the softness dependence of crystals by the vertical translation of the asymptotic lines of the hyperbolic operating curve. The solid asymptotic lines mean harder crystals than the dotted ones. (ii) Results of analysis of temperature and crystal softness dependence.

# Supplementary Material
# for
# Mathematical modeling of morphological changes in photochromic crystals by catastrophe theory

Hirotsugu Suzui[1], Kazuharu Uchiyama[2], Kingo Uchida[3], Ryoichi Horisaki[1], Hirokazu Hori[2] and Makoto Naruse[1]

[1] Department of Information Physics and Computing, Graduate School of Information Science and Technology, The University of Tokyo, 7-3-1 Hongo, Bunkyo-ku, Tokyo 113-8656, Japan.

[2] University of Yamanashi, 4-3-11 Takeda, Kofu, Yamanashi 400-8511, Japan.

[3] Department of Materials Chemistry, Ryukoku University, 1-5 Yokotani, Oecho, Seta, Otsu, Shiga 520-2194, Japan.

E-mail: tsuguh-hirosmtronica@g.ecc.u-tokyo.ac.jp

### Part 1: Crystal cracking after repeated irradiation of UV and visible light

The Supplementary Movie demonstrates the crystal bending and cracking due to UV irradiation. The experiment was performed at room temperature with the initial shape shown in Fig. S1(i). UV and visible light are alternately radiated from the left side. UV light irradiation induces gradual bending toward the right and sudden bending in the reverse direction, as described in the main manuscript. Visible light irradiation resets the crystal. However, after multiple repetition of UV and visible light irradiation, the accumulated mismatch of the crystal induces another type of morplogical change, which is cracking, as described in the main manuscript.

In the experiment, alternate irradiation with UV and visible light was conducted three times, followed by one additional UV light irradiation. In the movie, visible light irradiation was performed at the times 0:29, 2:08, and 6:05. Figure S1(ii) shows a snapshot during the second visible light irradiation. After the third visible light irradiation, the shape shown in Fig. S1(iii) was observed. After the bending induced by UV light irradiation as shown in Fig. S1(iv), the crystal suddenly cracks, as shown in Fig. S1(v).

To quantitatively evaluate the cracking, we analyzed the movie data after the time of 6:15 when the third visible light exposure was completed. In Fig. S2, the distance from the tip of the crack fragment to the crystal body with an angle of 45 degree with respect to the horizontal axis was evaluated as a function of the elapsed time from the starting time of the fourth UV light irradiation. We observe that cracking is suddenly induced at approximately 20 second.

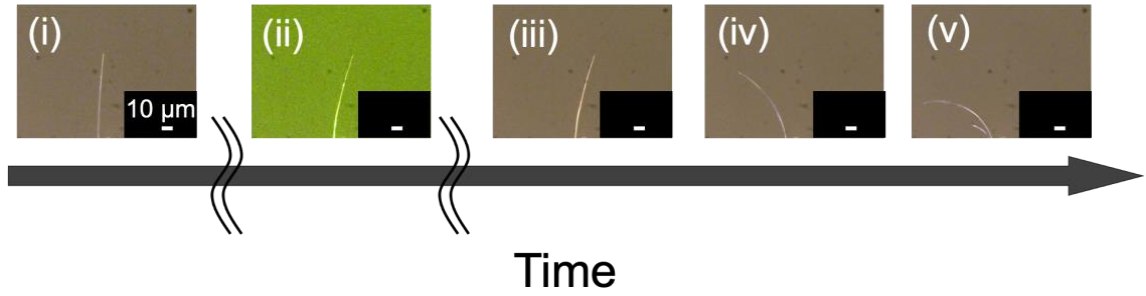

**Fig. S1** Crystal bending-and-cracking phenomena by alternate irradiation with UV and visible light.

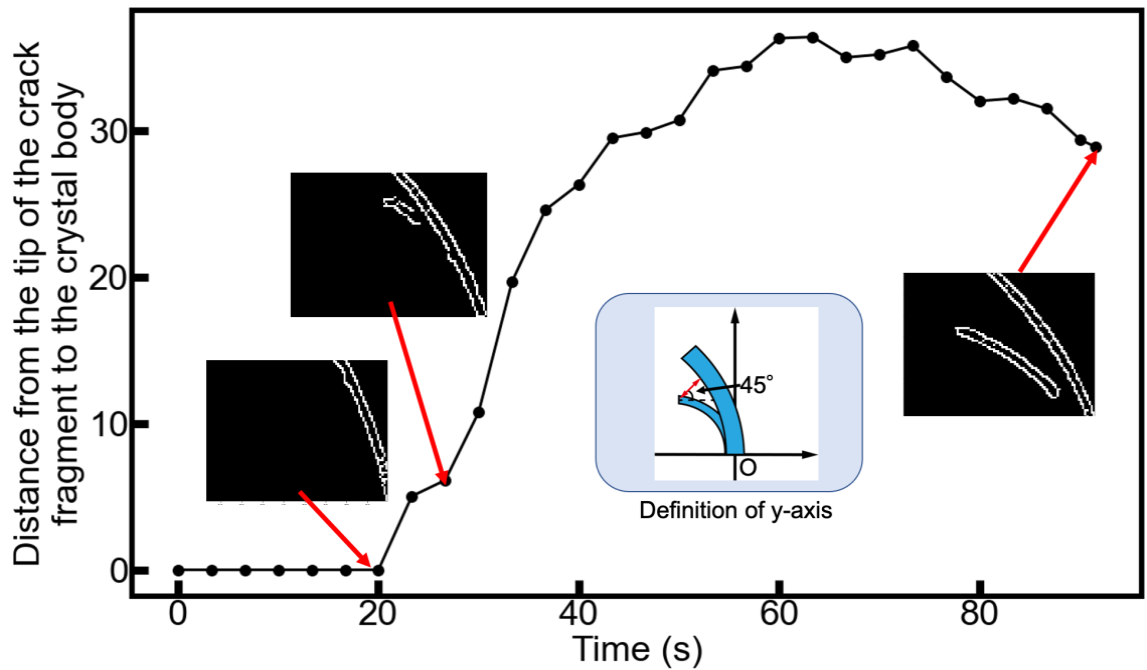

**Fig. S2** Image processing analysis of crystal bending-and-cracking phenomena. The distance from the tip of the crack to the crystal body with an angle of 45 degree with respect to the horizontal axis.

### Part 2: Theoretical derivations of the branch set in cusp catastrophe

The branch set in the cusp catastrophe, which is given by Eq. (4) in the main text, is derived by the following procedure.

The cusp singularity set is shown below, which is the same as Eq. (3) in the main text:

$$\Sigma_V = \left\{(u,v,x) \Big| \frac{\partial V}{\partial x} = x^3 + ux + v = 0, \frac{\partial^2 V}{\partial x^2} = 3x^2 + u = 0\right\}. \quad \text{(S1)}$$

The branch set $B_V$, which is the singularity set projected onto the $(u, v)$-plane, is expressed as

$$B_V = \left\{(u,v) \Big| \exists x, \begin{array}{ll} \frac{\partial V}{\partial x} = x^3 + ux + v = 0 & \text{(i)} \\ \frac{\partial^2 V}{\partial x^2} = 3x^2 + u = 0. & \text{(ii)} \end{array}\right\} \quad \text{(S2)}$$

We eliminate $x$ from these two equations (i) and (ii) in Eq. (S2). From (ii) of Eq. (S2),

$$u = -3x^2 \quad \text{(S3)}$$

holds. Substituting Eq. (S3) into (i) of Eq. (S2), we obtain

$$\left(-\frac{u}{3}\right)^3 = \left(\frac{v}{2}\right)^2. \quad \text{(S5)}$$

Hence,

$$B_V = \{(u,v) | 4u^3 + 27v^2 = 0\}. \quad \text{(S6)}$$

holds, which is Eq. (4) in the main text.

**Part 3: Theoretical derivations of the setting of the operating curve in the swallow-tail catastrophe modeling**

In Section III.B of the main text, the intersection of the asymptotic lines is given by $(v_0, w_0) = (0, -u^2/12)$. Here, we assume that the $w$-axis value of the origin coincides with that of the apex of the branch set. In considering the analysis after the introduction of the hyperbolic operating curve, the catastrophe is shifted so that the intersection of the asymptotes of the hyperbola and the height of the two cusp points of the catastrophe coincide. The parametric expression for the swallow tail catastrophe is given below.[1]

$$v = 2ut - 4ut^3, \quad \text{(S7)}$$

$$w = ut^2 - 3t^4. \quad \text{(S8)}$$

Differentiating Eqs. (S7) and (S8), respectively, we obtain

$$\frac{dv}{dt} = 2u - 12ut^2 \quad \text{(S9)}$$

$$\frac{dw}{dt} = 2ut - 12t^3. \quad \text{(S10)}$$

From Eq. (S10), the catastrophe takes extreme values at

$$t = -\sqrt{\frac{u}{6}}, 0, \sqrt{\frac{u}{6}}. \tag{S11}$$

The cusp points of the swallow-tail catastrophe, which gives the maximum of $w$, are obtained as

$$t = -\sqrt{\frac{u}{6}}, \sqrt{\frac{u}{6}}.$$

The value of $w$ at these $t$ values is given by Eq. (S8):

$$w = u\left(\sqrt{\frac{u}{6}}\right)^2 - 3\left(\sqrt{\frac{u}{6}}\right)^4 = \frac{u^2}{12}. \tag{S12}$$

Therefore, shifting the asymptotic lines by $w_0 = -\frac{u^2}{12}$ along the $w$-axis makes the $w$ value of their intersection the same as that of the cusp point given by Eq. (S12). This is the derivation of the intersection of the asymptotic lines given by $(v_0, w_0) = (0, -u^2/12)$.

**Reference**

1. David H. von Seggern, *CRC Standard Curves and Surfaces with Mathematica Second Edition.* (FL: CRC Press, 2007.)